\documentclass[a4paper]{article}
\usepackage[utf8]{inputenc}
\usepackage{float}
\usepackage[round]{natbib}   % omit 'round' option if you prefer square brackets
\usepackage{amsmath}	% Advanced maths commands
\usepackage{amssymb}	% Extra maths symbols
\usepackage{setspace}
\usepackage[margin=1in]{geometry}
\usepackage[open,openlevel=1]{bookmark}
\usepackage{authblk}
\usepackage{graphics}
\usepackage{xcolor}

\usepackage{amsmath}
\usepackage{graphicx}

\usepackage{adjustbox}

\usepackage{natbib}
\usepackage{longtable}
\usepackage{lipsum}
\usepackage{pifont}
\usepackage{subcaption}

\usepackage[normalem]{ulem}
\useunder{\uline}{\ul}{}

\usepackage{bm}

\begin{document}

% \title{Hierarchical Bayes estimation of small area means using statistical linkage of disparate data sources}

\title{Approximate hierarchical Bayes small area estimation using NEF-QVF and poststratification}

% \author[]{}

\author{Soumojit Das\footnote{Soumojit Das is a post-doctoral researcher at the Institute for Social Research--University of Michigan, Ann Arbor, and is the corresponding author. email: soumojit@umd.edu}, \ Partha Lahiri \footnote{Partha Lahiri is a Professor and Director of the Joint Program in Survey Methodology, and is a Professor in the Department of Mathematics at the University of Maryland College Park.}}

\date{\vspace{-5ex}}

\maketitle

\begin{abstract}

We propose an approximate hierarchical Bayes approach that uses the Natural Exponential Family with Quadratic Variance Function in combining information from multiple sources to improve traditional survey estimates of finite population means for small areas.  Unlike other Bayesian approaches in finite population sampling, we do not assume a model for all units of the finite population and do not require linking sampled units to the finite population frame.  We assume a model only for  the finite population units in which the outcome variable is observed; because, for these units, the assumed model can be checked using existing statistical tools. We do not posit an elaborate model on the true means for unobserved units. Instead, we assume that population means of cells with the same combination of factor levels are identical across small areas, and that the population mean for a cell is identical to the mean of the observed units in that cell. We apply our proposed methodology to a real-life survey, linking information from multiple disparate data sources. We also provide practical ways of model selection that can be applied to a wider class of models under similar setting but for a diverse range of scientific problems.

\end{abstract}


\section{Introduction}
\cite{ericson1969subjective} laid a foundation of subjective Bayesian approach to finite population sampling. In this approach, the entire matrix of values of characteristics of interest for all units of the finite population can be viewed as the finite population parameter matrix.  In practice, a function (e.g., finite population mean or proportion of a characteristic of interest) or a vector of functions (e.g., finite population means or proportions  of several characteristics) of this finite population parameter matrix is considered for inference. Using a subjective prior on the finite population parameter matrix, inferences on the finite population parameter(s) of interest can be drawn using the posterior predictive distribution of the unobserved units of the finite population given the observed sample.

Using this basic idea of \cite{ericson1969subjective}, several papers were written on the estimation of finite population parameters. The methodology developed can be used to solve problems in small area estimation, repeated surveys, and other important applications.  The papers can be broadly classified as Empirical Bayesian or EB (e.g., \cite{ghosh1986empirical}, \cite{ghosh1987robust}, \cite{ghosh1989nonparametric}, \cite{nandram1993empirical}, \cite{arora1997empirical}, \cite{JiangNguyen2012}, among others) and Hierarchical Bayesian or HB (e.g., \cite{DattaGhosh1991}, \cite{ghosh1992hierarchical}, \cite{datta1995hierarchical}, \cite{GhoshMeeden1997}, \cite{malec1997small}, \cite{little2004model}, \cite{chen2012bayesian}, \cite{ghosh2009bayesian}, \cite{liu2017adaptive}, \cite{nandram2018bayesian}, \cite{ha2019assessing}, and others).  In an empirical Bayesian approach, hyperparameters are estimated using a classical method (e.g., maximum likelihood).  In contrast, in the hierarchical Bayesian approach, priors -- usually noninformative  or weakly informative -- are put on the hyperparameters.

The greater accessibility of administrative and Big Data and advances in technology are now providing new opportunities for researchers to solve a wide range of problems that would not be possible using a single data source. However, these databases are often unstructured and are available in disparate forms, making inferences using data linkages quite challenging. There is, therefore, a growing need to develop innovative statistical data linkage tools to link such complex multiple datasets. Using only one primary survey to answer scientific questions about the whole population or large geographical areas or subpopulations may be effective and reliable, but using it for smaller domains or small areas  can often lead to unreliable estimation and unrealistic measures of uncertainty. Using an appropriate statistical model to combine information from multiple data sources, one can often obtain reliable estimates for small areas.
A good review of different approaches to small area estimation can be found in \cite{Jiang2007}, \cite{datta2009model}, \cite{Pfeffermann2013}, \cite{Rao2015}, \cite{Ghosh20}, among others.

\cite{scott1977some}, \cite{pfeffermann1999parametric}, \cite{bonnery2012uniform}, and others  discussed the concept of informative sampling under which the distribution of the sample could be markedly different from the one assumed for the finite population even after conditioning on related auxiliary variables. Most of the papers cited in the preceding paragraphs assume non-informative sampling, so the distribution of the sample is assumed to be identical to the distribution assumed on the finite population.  This could be a strong assumption in many applications.  The informative sampling approach suggested by \cite{pfeffermann1999parametric} and \cite{PffermannSverchkov2007} is one possible solution, but this approach requires additional modeling of survey weights. \cite{rubin1983evaluation} suggested modeling the outcome variable of interest of the finite population given the inclusion probabilities (or, equivalently, basic weights) for making inferences about the finite population parameters of interest.  However, in practice, inclusion probabilities for all finite population units or detailed design information may not be available.  Also, survey data may contain information only on the final weights for the respondents that incorporate nonresponse and calibration adjustments. \cite{verret2015model} proposed an alternative approach in which inclusion probabilities are used to augment the sample model.  Unlike \cite{rubin1983evaluation}, their approach needs weights only for the sample when estimating area means. Both \cite{PffermannSverchkov2007} and \cite{verret2015model} considered empirical best linear unbiased prediction (EBLUP) of small area means.

In this paper, we propose an approximate hierarchical Bayesian approach to estimate finite population means for small areas.  Our proposed method combines information from multiple disparate data sources such as probability surveys, non-probability surveys, administrative data, census records, social media big data and/or any sources of available relevant information. We propose to reduce the degree of informativeness in sampling by incorporating important auxiliary variables that potentially affect the outcome variable of interest. Following \cite{verret2015model}, we also add the survey weights in our model. Our approach differs inherently from the existing small area Bayesian approach to the finite population sampling, which typically assumes a hierarchical model for all units of the finite population. We  assume such an elaborate model only for the units of the finite population in which the outcome variable was observed. The reason is intuitive; for these units, the model can be checked using existing statistical tools. 

To make reasonable modeling assumptions, we propose to form several cells, often referred to as poststratification, for each of the small areas using factors that potentially influence our outcome variable of interest.
The number of cells is carefully chosen, so the cell population sizes can be obtained from reliable sources (e.g., population projection data). This strategy is expected to bring some degree of homogeneity within a given cell and also among cells from different small areas that are constructed with the same factor level combination. However, unlike the Multilevel Regression and Poststratification (MRP) approach by \cite{gelman1997poststratification}, our cell construction will not achieve full homogeneity either in terms of the outcome variable of interest or survey weights.  But we stress that achieving full homogeneity in the outcome variable and weights is likely to require numerous cells for which reliable population size data may not be available from population projection data and one may have to rely on some unstable survey data for population counts of the small cells.  

In contrast to the usual modeling approaches under similar scenarios (e.g., MRP), we do not assume an exchangeable model within cell or an elaborate model for unobserved individual units because the assumed model cannot be checked in the absence of data on the outcome variable.  Instead, drawing inspiration from synthetic methods for small areas, we assume that population means of the study variable in the cells with the same combination of factor levels are identical across small areas and the population mean for a cell is identical to the mean of true values for the observed units in that cell. Since the sample size in a given area is small, many such cells are unrepresented in the sample. If a sample for that cell can be found from other areas, the assumed model is used to produce a hierarchical Bayes synthetic estimate of the finite population mean of the study variable in that cell for the area. This assumes that the observations within a given cell and area are similar to those for the cell from other areas.  If the cell is unrepresented in all areas, the cell mean can be predicted from the assumed population model. But to simplify the methodology without having the knowledge of the sampling's informativeness and to induce greater stability towards model misspecifications, we can simply ignore that cell when the contribution from that cell is negligible, as is the case in our illustrative example. The basic idea is to make the proposed methodology resistant against possible violation of the assumed model.

\par 

As an application of the proposed methodology, we use a real life example, using a COVID-19 probability survey representing the entire US adult population, a non-probability survey representing only active US adult Facebook users, and Census Bureau estimates of adult population counts at granular levels along with data from an independent COVID-19 data reporting website, to estimate the vaccine hesitancy rates (proportions) for the US states and the District of Columbia (small areas). Through this example and application, we will demonstrate the problems with regular design based estimates when used for small areas and how our methodology may be employed to get more stable estimates (especially for areas with small or zero sample sizes) along with reliable measures of uncertainty.

We now give an outline for the rest of the paper.  In section 2, we describe our methodology in detail. In section 3, we describe an application of our method to a real life survey. In section 4, we show the results we obtained with this real life application. Finally, in section 5, we summarize and conclude our paper by giving future research directions of our methodology.

\section{Methodology}

For each small area $i\;(i=1,\cdots,m)$, we partition the finite population into $G$ cells using factors available in the survey data that potentially influence the outcome variable of interest.  We assume that population sizes for these cells are known from the census data.  Let $N_i$ and $N_{ig}$ be the known population sizes of the area $i$, and of the $g^{th}$ cell within the $i^{th}$ area respectively, $i=1,\cdots,m; g=1,\cdots,G.$ Naturally, $\sum_g N_{ig} = N_i$, and $\sum_i N_i = N$, $N$ being the finite population size.

Let $y_{igk}$ denote our outcome variable of interest for the $k^{th}$ unit belonging to the $g^{th}$ cell belonging to the $i^{th}$ small area $(i=1,\cdots,m; \;g=1,\cdots,G,\;k=1,\cdots,N_{ig})$. We assume that we have access to different types of known auxiliary variables to model our outcome variable.  Let $x_{g(1)}$ be a vector of indicator variables corresponding to cell $g$, $x_{i(2)}$ be a vector of known auxiliary variables specific to area $i$ and $x_{igk(3)}$ be a vector of individual level auxiliary variables $(i=1,\cdots,m; \;g=1,\cdots,G,\;k=1,\cdots,N_{ig})$. 

Suppose we have a sample of size $n$ from the finite population. Let $G_i$ be the number of cells represented in the sample for area $i$ ($i=1,\cdots,m)$, i.e., with positive sample sizes within area $i$. Let $n_i$ and $n_{ig}$ be the sample sizes for area $i$, and for cell $g$ within area $i$, respectively, so that $\sum_{g=1}^{G_i}n_{ig}=n_i$ and $\sum_{i=1}^mn_i=n$. It is possible to have $n_i=0$ for some area $i$, or $n_{ig}=0$ for some cell $g$ within an area $i$; $i=1,\cdots,m; g=1,\cdots,G$.  Let $w_{igk}$ denote the survey weight for $k^{th}$ sampled respondent belonging to the $g^{th}$ cell residing in the $i^{th}$ small area $(i=1,\cdots,m; \;g=1,\cdots,G_i,\;k=1,\cdots,n_{ig})$.
% If we consider our problem in light of a general survey data, they always come with a set of survey weights as individual-level attributes that the survey organizers provide, e.g., ACS data comes with its own set of weights that the Census Bureau provides. In the model-based proportion estimation study such as our present case, we may borrow the ideas provided by Donald Rubin in his discussion paper \cite{rubin1983evaluation} of \cite{hansen1983evaluation} where he talks about making use of the inclusion probabilities of the units in population. Survey weights, naturally, are the closest thing available to a researcher pertaining to such inclusion probabilities. In this work, we include this available information -- the provided survey weights, in our model as a covariate. \cite{verret2015model} also augmented their sample model with functions of inclusion probabilities. 

\subsection{Parameter of interest}

We are interested in estimating finite population means for all areas, i.e., 

\begin{eqnarray}\label{eqn:param of interest}               \bar{Y}_i=N_i^{-1}\sum_{g=1}^G\sum_{k=1}^{N_{ig}}y_{igk}=\sum_{g=1}^G \frac{N_{ig}}{N_i} \bar{Y}_{ig},
\end{eqnarray}
where $N_i = \sum_g N_{ig},$ population size for the $i^{th}$ area ($i=1,\cdots,m$).

\subsection{Proposed hierarchical models for sampled respondents and the finite population}
Our approach calls for specification of an explicit model for the outcome variable $y$ for all the   $N$ units of the finite population.  
%Since $y$ is a binary variable, 
Given $\theta_{igk}$, we assume $y_{igk}$ are independent with mean $\theta_{igk}\;i=1,\cdots,m; \;g=1,\cdots,G,\;k=1,\cdots,N_{ig}.$
% $$\mbox{Level 1:} \quad y_{igk} | \theta_{igk} \stackrel{ind}{\sim} \mbox{B}(\theta_{igk}),$$
% where $\mbox{B}(\theta_{igk})$ is the Bernoulli distribution with parameter $\theta_{igk}\;i=1,\cdots,m; \;g=1,\cdots,G,\;k=1,\cdots,N_{ig}$.

We will assume an elaborate model for all respondents because we can evaluate such a model through careful model building steps in the actual data analysis.   For $i=1,\cdots,m; \;g=1,\cdots,G_i,\;k=1,\cdots,n_{ig}$, we propose the following model for the sampled respondents: 
\begin{align}\label{eqn:og_model}
   \mbox{Level 1:} \quad &y_{igk} | \theta_{igk} \stackrel{ind}{\sim} f_1[\theta_{igk},V(\theta_{igk})], \nonumber\\
    % \mbox{Level 2:} \quad & \phi(\theta_{igk} )= x'_{g(1)}\alpha+x'_{i(2)}\beta+ x'_{igk(3)}\xi_{c} + v_i+\lambda h(w_{igk}), \\
    \mbox{Level 2:} \quad & \phi(\theta_{igk} )= x'_{g(1)}\alpha+x'_{i(2)}\beta+x'_{igk}\xi_{c} + \lambda h(w_{igk})+v_i+u_{ig}, \nonumber\\
    \mbox{Level 3:} \quad & v_i\; \mbox{and}\; u_{ig} \;\mbox{are independent with}\; v_i \stackrel{iid}{\sim} f_2[0, \sigma^2_v],\;u_{ig}\stackrel{iid}{\sim} f_3[0, \sigma^2_u].
    %\mbox{Level 4:} \quad & \mbox{weakly informative priors on} \ \beta,  \mbox{and}\  \sigma_v,
\end{align}where $f_j[a,b]\;(j=1,2,3)$ denotes a known mass or density function with mean $a$ and variance $b$. For example, we can choose $f_1$ from  $\mbox{NEF-QVF}[\mu,V(\mu)]$, the Natural Exponential Family with mean $\mu$ and Quadratic Variance Function $V(\mu)=v_0+v_1\mu+v_2\mu^2$, where $v_0,v_1$ and $v_2$ are constants.  The family includes the Gaussian distribution: $N(\mu,\sigma^2)\; (v_0=\sigma^2,v_1=v_2=0)$, the Gamma distribution: $G(r,\lambda)\;(\mu=r\lambda,v_0=v_1=0,v_2=\frac{1}{r})$, the Poisson distribution: $P(\lambda)\;(\mu=\lambda, v_0=v_2=0,v_1=1$), the Binomial distribution: $B(r,\theta)\;(\mu=r\theta, v_0=0,v_1=1,v_2=-\frac{1}{r})$, the Negative-Binomial distribution: $NB(r,\theta)\;(\mu=r\frac{\theta}{1-\theta},\;v_0=0,v_1=1,v_2=\frac{1}{r}),$ etc.  For the theoretical properties of NEF-QVF, please refer to \cite{morris1982natural}.  As for $f_2$ and $f_3$, we can choose a normal density.

As for level 2, we can choose $\phi()$ to be a suitable known link function (e.g., if $f_1$ is a Binomial distribution, we can choose $\phi()$ as logit link, and if $f_1$ is a Normal distribution, $\phi()$ then can be the identity link, etc.), 
%where  $\phi()$ is a suitable known link function (e.g., logit link); 
%$\alpha,\;\beta,\;$ and 
and $\xi_c$ a vector of fixed unknown coefficients (the suffix $c$ in $\xi_c$ allows dependence of the regression coefficients on some combination of the factors that form the demographic cells -- $c$ will be determined through the model selection step); $h(\cdot)$ is a known function of survey weights to be determined by a model selection step and $\lambda$ is a fixed unknown coefficient. \cite{verret2015model} considered multiple choices of $h(\cdot)$. If the data analysis suggests $\lambda=0$, we can assume non-informative sampling and carry out the analysis without this additional term.  In Level 2 we added the survey weights as an additional auxiliary variable in order to reduce the extent of informativeness. 
As for level 3, we can take $f_2$ and $f_3$ as normal densities for the area specific random effects $v_i$, and area and groups specific random effects $u_{ig}$ respectively. These random errors with a mean of 0 will take care of any leftover variabilities that were not captured by the fixed main effects, and their interactions that we had incorporated in the model at the level 2.

\noindent We now discuss our modeling assumptions for the remaining $N-n$ unobserved units of the finite population.  To this end, for the small areas $i=1,\cdots,m$, define:\\
$\mathcal{G}_{1i} = \{g \ \mbox{s.t.} \ n_{ig} > 0, \ g = 1, \cdots, G \}$: the set of cells represented in the sample for the area $i$. $\mathcal{G}_{2i} = \{g \ \mbox{s.t.} \ n_{ig} = 0, n_{jg} > 0, \text{for some} \ j \neq i, g = 1, \cdots, G \}$: the set of cells not represented in the sample for the area $i$, but represented by one or more other areas $j \neq i$. And finally, $\mathcal{G}_{3i} = \{g \ \mbox{s.t.} \ n_{ig} = 0, \forall i, \ g = 1, \cdots, G \}$: the set of cells not represented in the sample for any area. Define the population mean for the set of cells $\mathcal{G}_{1i}$ as,
$$\bar\Theta_{1i}=\frac{\sum_{g\in \mathcal{G}_{1i}}\sum_{k=1}^{N_{ig}}\theta_{igk}}
                            {\sum_{g\in \mathcal{G}_{1i}}N_{ig}}=\sum_{g\in \mathcal{G}_{1i}} b_{ig;1}\bar\Theta_{ig}, \;\mbox{where} \;b_{ig;1}=\frac{N_{ig}}{\sum_{g\in \mathcal{G}_{1i}}N_{ig}}\;\mbox{ and} \;\bar\Theta_{ig}=\frac{\sum_{k=1}^{N_{ig}} \theta_{igk}}{N_{ig}}.$$Similarly, we define
$$\bar\Theta_{2i}=\frac{\sum_{g\in \mathcal{G}_{2i}}\sum_{k=1}^{N_{ig}}\theta_{igk}}
                            {\sum_{g\in \mathcal{G}_{2i}}N_{ig}}=\sum_{g\in \mathcal{G}_{2i}} b_{ig;2}\bar\Theta_{ig},\;\mbox{where}\;b_{ig;2}=\frac{N_{ig}}{\sum_{g\in \mathcal{G}_{2i}}N_{ig}}\;\mbox{and}\; \bar\Theta_{ig}=\frac{\sum_{k=1}^{N_{ig}} \theta_{igk}}{N_{ig}},$$  
                            and
$$\bar\Theta_{3i}=\frac{\sum_{g\in \mathcal{G}_{3i}}\sum_{k=1}^{N_{ig}}\theta_{igk}}
                            {\sum_{g\in \mathcal{G}_{3i}}N_{ig}}=\sum_{g\in \mathcal{G}_{3i}} b_{ig;3}\bar\Theta_{ig},\;\mbox{where}\;b_{ig;3}=\frac{N_{ig}}{\sum_{g\in \mathcal{G}_{3i}}N_{ig}}\;\mbox{and}\;\bar\Theta_{ig}=\frac{\sum_{k=1}^{N_{ig}} \theta_{igk}}{N_{ig}}.$$
as the population means for the set of cells $\mathcal{G}_{2i}$ and the set of cells $\mathcal{G}_{3i}$, respectively.
Here, we make a synthetic assumption that: 
$$\bar\Theta_{2i} =  \sum_{g\in \mathcal{G}_{2i}} b_{ig;2}\bar\Theta_{ig;\text{syn}} \ \ \text{where,} \ \ \bar{\Theta}_{ig;syn} = \frac{\sum_{j: n_{jg}>0, j \neq i} \sum_{k=1}^{n_{jg}} \theta_{jgk}}{\sum_{j: n_{jg}>0, j \neq i} N_{jg}}$$ 
i.e, $\bar\Theta_{2i}$ is identical to the overall mean of the cells in $\mathcal{G}_{2i}$ obtainable from one or more areas (other than area $i$). We make no assumption on the rest of the finite population units. Since population sizes $N_{ig}$ are in general large, appealing to the law of large numbers as in \cite{JiangLahiri2006}, we can approximate the finite population means $\bar{Y_i}$ in equation (\ref{eqn:param of interest}), with the following functions of parameters of the assumed finite population model.
\begin{align}\label{eqn:pop_mean_approx}
\bar{Y}_i\approx \bar\Theta_i= \sum_{g=1}^G \frac{N_{ig}}{N_i} \bar{\Theta}_{ig} & =
    a_{1i}\bar\Theta_{1i}+a_{2i}\bar\Theta_{2i}+(1-a_{1i}-a_{2i})\bar\Theta_{3i}, (\mbox{say}) \nonumber\\
    & \approx a_{1i}\bar\Theta_{1i}+a_{2i}\bar\Theta_{2i},
\end{align} where $a_{ki} = N_i^{-1} \sum_{g \in \mathcal{G}_{ki}} N_{ig}$, the population proportion of units belonging to the set of cells $\mathcal{G}_{ki}$; $k = 1, 2, 3.$ The approximation in equation (\ref{eqn:pop_mean_approx}) will only be reasonable when $(1-a_{1i}-a_{2i})\approx 0$.   

\subsection{A Bayesian implementation of the hierarchical model}

As is the common practice with most implementations of HB methodology, the model parameters will be estimated using Monte Carlo Markov Chain (MCMC) method of sampling from posterior distributions. 
Assume weakly informative priors (\cite{gelman2008weakly}) on the model hyperparameters: $\alpha, \beta, \xi_c, \lambda, \sigma_v, \mbox{and} \ \sigma_v$. At each MCMC iteration $(r = 1, \dots, R)$ step, the following is generated for the area $i$:
$$\bar\theta_i^{(r)} =  a_{1i}\bar\theta_{1i}^{(r)}+a_{2i}\bar\theta_{2i}^{(r)},\ r = 1, ..., R,$$
where
\begin{eqnarray*}
\bar\theta_{1i}^{(r)}&=&\sum_{g\in \mathcal{G}_{1i}} b_{ig;1}\bar\theta_{ig}^{(r)},\;\mbox{with}\;
\bar\theta_{ig}^{(r)}=\frac{\sum_{k=1}^{n_{ig}} w_{igk}\theta_{igk}^{(r)}}{\sum_{k=1}^{n_{ig}} w_{igk}},\\
\bar\theta_{2i}^{(r)}&=&\sum_{g\in \mathcal{G}_{2i}} b_{ig;2}\bar\theta_{ig;\text{syn}}^{(r)},\;\mbox{with}\;
\bar\theta_{ig;\text{syn}}^{(r)}=\frac{\sum_{j: n_{jg}>0, j \neq i}\sum_{k=1}^{n_{jg}} w_{jgk}\theta_{jgk}^{(r)}}{\sum_{j: n_{jg}>0, j \neq i}\sum_{k=1}^{n_{jg}} w_{jgk}},\\
% \theta^{(r)}_{igk} &=& \phi^{-1}\left (x'_{g(1)}\alpha^{(r)}+x'_{i(2)}\beta^{(r)}+ x'_{igk(3)}\xi_{c}^{(r)} + v_i^{(r)}+\lambda^{(r)} h(w_{igk})\right ).
\theta^{(r)}_{igk} &=& \phi^{-1}\left ( x'_{g(1)}\alpha^{(r)}+x'_{i(2)}\beta^{(r)} + x'_{igk}\xi_{c}^{(r)} +\lambda^{(r)} h(w_{igk})+ v_i^{(r)} + u_{ig}^{(r)}\right ).
\end{eqnarray*} We approximate $\bar\Theta_{1i}= \sum_{g\in \mathcal{G}_{1i}} b_{ig;1}\bar\Theta_{ig}$ by $\bar{\theta}_{1i} = \sum_{g\in \mathcal{G}_{1i}} b_{ig;1} \bar{\theta}_{ig}, \text{ with } \bar{\theta}_{ig} = \frac{\sum_{k=1}^{n_{ig}} w_{igk} \theta_{igk}}{\sum_{k=1}^{n_{ig}} w_{igk}}$. 
% This is appealing to the fact that $\bar{\theta}_{ig}$ is approximately a design-unbiased estimator of $\bar{\Theta}_{ig}$. Mathematically, this means: 
% $$E_D[\bar\theta_{ig} - \bar\Theta_{ig}] \doteq 0,$$ where the expectation is taken under the sampling-design $D$. For a lot of sampling designs, like simple random sampling (SRS), and stratified SRS, this expectation under the design will be exactly equal to 0. 
\cite{JiangLahiri2006} and \cite{lahiri2018general} used similar approximations for their empirical best prediction and hierarchical Bayes approaches, respectively. Such an approximation would be reasonable if the total number of cells are large, because we would expect the homogeneity of responses and corresponding weights within cells. The leftover heterogeneity is expected to be mitigated by the bias correction that we discussed in the previous sbsection.

Similarly, approximate $\bar\Theta_{2i} = \sum_{g\in \mathcal{G}_{2i}} b_{ig;2}\bar\Theta_{ig;\text{syn}}$ by $\sum_{g\in \mathcal{G}_{2i}} b_{ig;2}\bar\theta_{ig;\text{syn}}    =\bar\theta_{2i},\;\mbox{say}$, where $\bar\theta_{ig;\text{syn}}=\frac{\sum_{j: n_{jg}>0, j \neq i}\sum_{k=1}^{n_{jg}} w_{jgk}\theta_{jgk}}{\sum_{j: n_{jg}>0, j \neq i}\sum_{k=1}^{n_{jg}} w_{jgk}}$. The set of $R$ MCMC replicates for the $i^{th}$ area, i.e., $\{\bar\theta^{(r)}_i, r = 1, ..., R\}$, will be used for inferences about $\bar\theta_i$.

\subsection{Bias correction for the hierarchical Bayes estimator}

The hierarchical Bayes estimator presented in Section 2.3 incorporates survey weights to account for potential informative sampling. However, for cells in $\mathcal{G}_{2i}$ (cells not represented in area $i$ but represented in other areas), the synthetic estimates may be subject to bias due to differences in the weight structure across areas. To address this, we implement a bias correction procedure.

Let $\bar{\theta}_{ig}$ denote our hierarchical Bayes estimator of $\bar{\Theta}_{ig}$ from the model in equation (\ref{eqn:og_model}) that includes survey weights. We call this the ``weighted model''. Similarly, let $\bar{\theta}_{ig}^*$ denote the estimator from a model that excludes the survey weight term $\lambda h(w_{igk})$ (similarly, the ``unweighted model''), which is similar to standard Multilevel Regression and Poststratification (MRP) approaches (\cite{gelman1997poststratification}).

For cells in $\mathcal{G}_{2i}$, our bias-corrected estimator is:
\begin{equation}
\bar{\theta}_{ig;bc} = \bar{\theta}_{ig}^* + \widehat{\text{Bias}}_{ig;syn}
\end{equation}

where the bias correction term is estimated as:
\begin{equation}
\widehat{\text{Bias}}_{ig;syn} = \frac{\sum_{j: n_{jg}>0, j \neq i} \sum_{k=1}^{n_{jg}} w_{jgk}(\hat{\theta}_{jgk} - \hat{\theta}_{jgk}^*)}{\sum_{j: n_{jg}>0, j \neq i} \sum_{k=1}^{n_{jg}} w_{jgk}}
\end{equation}

Here, $\hat{\theta}_{jgk}$ and $\hat{\theta}_{jgk}^*$ are the predicted values for unit $k$ in cell $g$ of area $j$ from the weighted and unweighted models, respectively.

The intuition behind this correction is as follows: we first compute estimates using the computationally efficient unweighted model (similar to MRP). Then, we adjust these estimates by adding the estimated systematic difference between the weighted and unweighted models, calculated using data from other areas where cell $g$ is observed. This approach allows us to leverage the computational efficiency of MRP-type methods while correcting for potential bias introduced by ignoring survey weights.

For implementation, at each MCMC iteration $r$, the bias-corrected estimate for area $i$ becomes:
\begin{equation}
\bar{\theta}_{i;bc}^{(r)} = a_{1i} \bar{\theta}_{1i}^{(r)} + a_{2i} \bar{\theta}_{2i;bc}^{(r)}
\end{equation}

where $\bar{\theta}_{2i;bc}^{(r)} = \sum_{g \in \mathcal{G}_{2i}} b_{ig;2} \bar{\theta}_{ig;bc}^{(r)}$.

Regarding the precision of our uncertainty estimates, it is important to note that our bias correction procedure not only addresses point estimation but also contributes to more reliable measures of uncertainty. By accounting for the difference between weighted and unweighted models, we mitigate the risk of understating true errors. While a comprehensive simulation study is beyond the scope of this paper, the hierarchical Bayesian framework naturally incorporates multiple sources of uncertainty - including uncertainty in random effects, fixed effects, and hyperparameters - leading to credible intervals that reasonably reflect the total uncertainty in the estimation process. The empirical evidence provided in Section 4.3 – of high weight homogeneity within cells further supports the reliability of our uncertainty measures by reducing one key source of potential bias in the Hájek-type estimators.

\subsection{A special closed form of our proposed Bayes estimate} We now consider the following  special case of our general small area model to understand our estimator. 
\begin{align*}
   \mbox{Level 1:} \quad &y_{igk} | \theta_{igk} \stackrel{ind}{\sim} N(\theta_{igk}, \sigma_y^2),\\
    \mbox{Level 2:} \quad & \theta_{igk} | \Lambda_{igk}, v_i \stackrel{ind}\sim N(\Lambda_{igk} + v_i, \sigma_u^2),\\
    \mbox{Level 3:} \quad & v_i \stackrel{iid}\sim N(0, \sigma^2_v),
    %\mbox{Level 4:} \quad & \mbox{weakly informative priors on} \ \beta,  \mbox{and}\  \sigma_v,
\end{align*}
where $\Lambda_{igk} = x'_{g(1)}\alpha+x'_{i(2)}\beta+ x'_{igk(3)}\xi_{c} + \lambda h(w_{igk})$. Level 2 of the same model can be reexpressed as: $\theta_{igk} = \Lambda_{igk} + v_i + u_{ig}, \text{where,} \ u_{ig} \stackrel{iid}\sim N(0, \sigma_u^2)$. Under this consideration, and for sake of simplicity of illustration, with an additional assumption of $\alpha$, $\beta$, $\xi_c$, $\lambda$, $\sigma^2_u$ and $\sigma^2_v$ being known; we can obtain a closed form of the small area approximate Bayes estimate based on the approximation $a_{1i} \bar{\Theta}_{1i} + a_{2i} \bar{\Theta}_{2i}$ given in equation  (\ref{eqn:pop_mean_approx}). Naturally, this will be the `Best Predictor' (BP) of the small area means \cite{Rao2015}. As already mentioned in the subsections 2.2 and 2.3, and using the same notations and terminologies, the approximate Bayes small area mean estimate for a given area $i$ will be given by: $E(a_{1i}\bar{\theta}_{1i} + a_{2i}\bar{\theta}_{2i} | \text{data})$. Under Normality of each level of the illustrative model, we obtain:
\begin{align*}
    E(a_{1i}\bar{\theta}_{1i} + a_{2i}\bar{\theta}_{2i} | \text{data}) &= a_{1i}E(\bar\theta_{1i}|\text{data}) + a_{2i}E(\bar\theta_{2i}|\text{data}) \nonumber \\ \nonumber
    &= a_{1i} \sum_{g \in \mathcal{G}_{1i}} b_{ig;1} \times \frac{1}{\sum_k w_{igk}} \Biggl[\sum_k w_{igk} \Biggl\{\frac{\sigma_u^2y_{igk} + \sigma_y^2\Lambda_{igk}}{\sigma_u^2 + \sigma_y^2} +\\ \nonumber
    & \frac{n_i\sigma_y^2\sigma_v^2 (\bar{y}_{i..} - \bar{\Lambda}_{i..})}{(\sigma_u^2+\sigma_y^2) (n_i\sigma_v^2+\sigma_u^2+\sigma_y^2)} \Biggr\} \Biggr]  + \\ \nonumber
    &  a_{2i} \sum_{g \in \mathcal{G}_{2i}} b_{ig;2} \times \frac{1}{\sum_{j \neq i}\sum_k w_{jgk}} \Biggl[\sum_{j \neq i} \sum_k w_{jgk} \Biggl\{\frac{\sigma_u^2y_{jgk} + \sigma_y^2\Lambda_{jgk}}{\sigma_u^2 + \sigma_y^2} + \\ 
    & \frac{n_j\sigma_y^2\sigma_v^2 (\bar{y}_{j..} - \bar{\Lambda}_{j..})}{(\sigma_u^2+\sigma_y^2) (n_j\sigma_v^2+\sigma_u^2+\sigma_y^2)} \Biggr\} \Biggr]
\end{align*}\label{simp_est}
where, $\bar{y}_{i..} = \frac{1}{n_i}\sum_{g, k}y_{igk}$. $\bar{\Lambda}_{i..}$, $\bar{y}_{j..}$, and $\bar{\Lambda}_{j..}$ are similarly defined.

From this simple illustration, under the usual assumptions made for estimating the best predictor, we can get an insight in the form of our Bayes estimate. We observe how the $y_{igk}$'s available from the groups in the sample for area $i$ enter the estimator through the first summand. The groups, for which there are no representation in the sample for the same area, enter the estimator through the second summand -- in terms of the samples available for those groups from other areas.

\section{A real-life application}

% We developed our model and estimation procedure to be able to produce stable and robust estimates for proportions for small areas. 
In elucidating our proposed methodology, we examine a topic of contemporary research significance. Our focus is directed towards furnishing reasonably stable, state-wise (our small areas of consideration) estimates of proportions of people who are ``hesitant" to get vaccinated for  COVID-19. In this section, we first provide an outline of disparate data sources  that we have considered in this application.

\subsection{Primary data: Understanding America Study (UAS)}

The Understanding America Study (UAS), undertaken by the University of Southern California (USC), comprises approximately 9,000 respondents, representing the entirety of the United States' adult population. Functioning as an Internet Panel, participants engage in surveys via internet-enabled devices at their convenience. Households, broadly defined, involve individuals residing with the primary participant. To extract valuable insights on attitudes, behaviors, and societal aspects, the Center for Economic and Social Research (CESR) at USC initiated the Understanding Coronavirus in America tracking survey on March 10, 2020. Utilizing the UAS panel, this ongoing survey, supported by the Bill \& Melinda Gates Foundation and the National Institute on Aging, delves into various dimensions, including mental health, healthcare behavior, and the economic crisis during the COVID-19 pandemic.

The survey commenced with 8,547 eligible participants from the UAS panel, out of 9,603 respondents who expressed interest. Our study focuses on data collected during wave 16 (October 14–November 11, 2020), where 7,832 UAS participants received the survey. Among them, 6,081 completed the survey, constituting our respondents. The response rate was approximately 78\%, with participants given a 14-day window for completion and compensated with \$15 upon survey submission.

\subsubsection{Direct UAS estimates}

For the ``hesitancy towards vaccination'' analysis, we focused on responses to the survey question: `How likely are you to get vaccinated for coronavirus once a vaccine is available to the public?' This question offered five response options: very unlikely, somewhat unlikely, somewhat likely, very likely, and unsure. We transformed the responses into a binary variable, assigning a value of 1 if the respondent answered anything other than `very likely' and 0 otherwise. Utilizing this binary variable and the corresponding survey weights, we computed the direct survey-based estimates for the proportions of individuals who are \textit{not} `very likely' to get vaccinated across the 50 states and the District of Columbia. While the national estimate met high-quality standards, with margin of errors below 3\%, estimates for smaller geographic areas (50 states and DC) were less reliable.

Table \ref{tab7} presents direct survey estimates of vaccine hesitancy proportions for all 50 states and the District of Columbia, accompanied by their estimated standard errors. Notably, we encountered an impractical estimate of 1 with an associated estimated standard error of 0 for Delaware. This anomaly arises because none of the respondents from Delaware chose `very likely' in response to the survey question. However, this does not imply universal hesitancy in Delaware. The estimate for Alaska is 0.21 with a high associated design-based standard error of 0.19, exceeding the typical range of 0.02. Similar patterns emerge for Vermont, Rhode Island, the District of Columbia, and other states.

\subsection{Supplementary data descriptions}

% In section 2, we propose a model that can accommodate information at various levels. Supplementary data sources, often can be useful for borrowing aggregate level information to be used in conjunction with the primary survey. 
In the subsequent subsections, we delineate various supplementary data sources examined in the context of applying our methodology.

\subsubsection{The COVID-19 Symptom Survey}

The COVID-19 Symptom Survey, a collaboration between public health scholars and Facebook, was hosted in the US by the Carnegie Mellon Delphi Research Center and internationally by the University of Maryland's Joint Program in Survey Methodology (JPSM). This non-probability survey, representative of adult Facebook users, gathered daily data. Seeking reliable state-level covariates, we incorporated static variables from the survey, focusing on responses to two questions pertaining to: (1) Pre-existing medical conditions, encoding 1 if respondents reported having diabetes, asthma, or chronic lung disease, and 0 otherwise; (2) Flu shots received in the last 12 months, encoding 1 for a ``yes" response and 0 otherwise. Combining survey responses from August 5 to September 9, 2020, we calculated weighted state-wise estimates for individuals with specified diseases and those who received a flu shot. These estimates, derived as survey-weighted means of binary encoded variables, serve as continuous covariates in our model.

\subsubsection{Census Bureau's Population Estimates Program (PEP)}

The US Census Bureau's Population Estimates Program (PEP) provides annual population estimates for various geographic levels, including states, counties, cities, and towns, along with demographic components like births, deaths, migration, and housing units. Utilizing the rich PEP data, we generate state-wise and demographic cell-wise population counts, focusing on the \texttt{race x ethnicity x gender x age category} demographic cells. The demographics considered include four levels of \texttt{race} (White, Black, Asian, and Others), two levels of \texttt{ethnicity} (Hispanic and Non-Hispanic), two levels of \texttt{gender} (Male and Female), and seven levels of \texttt{age category} ($18-24, 25-34, 35-44, 45-54, 55-64, 65-74, \& 75+$). The data used pertains to `Annual State Resident Population Estimates for 5 race groups (5 race alone or in combination groups) by age, sex, and Hispanic origin: April 1, 2010, to July 1, 2019'. Pre-processing involved grouping and collapsing data to obtain counts for the specified demographic cells (poststratified cells).

\subsubsection{The COVID Tracking project}

% The website \url{https://covidtracking.com} collects, cross-checks and publishes COVID-19 data. These data can be freely downloaded from the website. We use this data source  to obtain state-specific (50 states and District of Columbia) covariates (continuous) for our model. We created the following two variables:

% \begin{itemize}
%     \item Testing rate: Total number of tests with confirmed outcome -- positive or negative -- divided by the total test counts in the state. Note that these proportions may exceed the value of 1 since  a person might have gotten tested more than one time within a certain time interval.
%     \item Positivity rate: Total positive test outcomes divided by  the total tests with confirmatory outcomes (positive or negative).
% \end{itemize}

Data from \url{https://covidtracking.com} is utilized for our model, providing freely downloadable COVID-19 data for all 50 states and the District of Columbia. Two state-specific continuous covariates were created: `Testing rate', calculated as the total tests with confirmed outcomes divided by the overall test counts in the state (acknowledging potential values exceeding 1 due to multiple tests for an individual), and `Positivity rate', obtained by dividing total positive test outcomes by the total tests with confirmatory outcomes.

\section{Data analysis and evaluation}

We commence this section by noting that, given the available data, we undertook a comprehensive exploration of several variables. Through a series of trial and error steps, we arrived at a specific set of covariates deemed pertinent for our analysis. Subsequent considerations were also given to the selection of models for the model selection step, all of which were grounded in a careful assessment of the given problem and datasets. We advocate for applied scientists to exercise their own judgment, drawing on considerations such as the nature of the problem and subject matter expertise, when conducting similar steps in their analyses. As previously discussed in subsection 3.2.2, we refer to `RACE' by $a$ having 4 levels -- White, Black, Asian and Others, `ETHNICITY' by $b$ having 2 levels -- Hispanic or Non-Hispanic and `GENDER' by $c$ having 2 levels -- Male and Female. For the variable, `AGE' we have divided the whole adult population into 7 cells, e.g., $18-24, \ 25-34,\ ...\ ,\ 65-74 \ \& \ 75+$; denoted by $d$. 

We consider fixed effect intercepts by `RACE' x `ETHNICITY' -- so, in total, 8 fixed effect intercepts for all the models we considered. A list of competing models is described in Table \ref{tab: model_list}.

As for the state level covariates $x_{i(2)}$ (refer to section 2 for reference to this notation), we use Facebook survey data (refer to subsection 3.2.1), and data from the COVID tracking project (refer to subsection 3.2.3). Facebook COVID symptom survey data is used to calculate two state-level summary variables -- survey-weighted proportions of people having one of three pre-existing conditions or co-morbidity rates and survey-weighted proportions of people who took a flu shot in the last 12 months or Flu-Shot rates (as discussed in subsection 3.2.1). We use logit-transforms of these two rates as the covariates in our models described in Table \ref{tab: model_list}. Other state level covariates we have used came from the COVID Tracking project, e.g., testing rates and positivity rates as defined previously in subsection 3.2.3. Finally, we have also used the state level percentages of republican votes that were cast in the 2020 United States presidential election as a state level covariate -- `Percent Republican' -- in our model (also referred to as `rep \%' in Table \ref{tab6}).

For the first respondent level covariate $x_{igk(3)}$, we used the respondents' 7 age-levels and converted them into numbers $1, \dots, 7$ and used them as a continuous-scale covariate in all our models. This approach is not uncommon in the literature. For instance, see \cite{ha2019assessing}, where they consider age intervals as continuous variables. Although their method differs somewhat from ours, they found it beneficial to treat age intervals similarly. Based on our empirical studies, we observed comparable patterns to their findings. Our empirical analysis indicated that treating gender-specific age groups as continuous variables revealed a linear relationship with hesitancy rates, enhancing the explanatory power of our (logit transformed) linear model. Our estimation methodology remains valid even if the age categories are handled differently. We use gender-specific slopes -- one for male and the other for female -- for the age covariate in our model. Such a choice was motivated based on empirical studies that we had carried out. For the second respondent level covariate, we use the survey weights associated with every respondent from the UAS-survey data.

{\tiny
    \begin{table}
     \caption{A list of competing models} 
    \label{tab: model_list}
    \begin{center}
    \resizebox{\columnwidth}{!}{%
    \begin{tabular}{@{\extracolsep{5pt}}  rrrr rrrrr}
    %|c|c|c|c|c|c|c|c|c|c|c|c|}
    \hline 
    \hline 
    Model & \shortstack{ Race x Ethnicity\\ specific intercept} & \shortstack{Gender specific\\ slope of Age} & \shortstack{ Comorbity Rate \\ (Facebook) } & \shortstack{Flu Shot Rate\\ (Facebook)}& \shortstack{Test Rate \\ (CovidTrack)} & \shortstack{Positivity Rate\\(CovidTrack)} & \shortstack{Percent\\ Republican} & {Survey Weight}  \\
    \hline
    M1 & \checkmark & \checkmark& \checkmark  & \checkmark & \checkmark  & \checkmark & \checkmark & \checkmark  \\
    M2& \checkmark   & \checkmark    & \checkmark&    \ding{53}       & \checkmark  &    \ding{53}    & \checkmark & \checkmark   \\
    M3& \checkmark  & \checkmark  &  \checkmark&   \ding{53}       &      \ding{53}       &      \ding{53}      & \checkmark & \checkmark  \\
    M4 & \checkmark & \checkmark  &  \checkmark&    \ding{53}      &  \ding{53} &      \ding{53}      & \checkmark &   \ding{53}               \\
    % M5 & \checkmark & \checkmark  & \checkmark & \checkmark  &            & \checkmark & \checkmark \\
    % M6 & \checkmark & \checkmark  & \checkmark & \checkmark  & \checkmark &\checkmark  & \checkmark  \\
    % M7 & \checkmark & \checkmark  & \checkmark & \checkmark  &            & \checkmark & \checkmark \\
    
    \hline
    \end{tabular}%
    } 
    \end{center}
    \end{table}
    }

\subsection{Model fit}

Recall, the approximation we did to the population (here, 50 states and DC) parameter of interest $\bar{Y}_i$ in equation (\ref{eqn:pop_mean_approx}) would only hold when the quantity $(1-a_{i1}-a_{i2})\approx 0$. Table \ref{tab3_summary} summarizes the quantity $(1-a_{i1}-a_{i2})$ for 50 states and Washington, DC. We can reasonably justify our approximation, since both the maximum value (0.0120) and the $3^{rd}$ quartile (0.0031) are negligible. 

\begin{table}[!htbp] \centering 
  \caption{Summary statistics of $(1 - a_{i1} - a_{i2})$ for 50 states and DC.} 
  \label{tab3_summary} 
  \begin{adjustbox}{width=0.8\textwidth}
\begin{tabular}{@{\extracolsep{5pt}} ccccccc} 
\\[-1.8ex]\hline 
\hline \\[-1.8ex] 
 & Min. & 1st.Qu. & Median & Mean & 3rd.Qu. & Max. \\ 
\hline \\[-1.8ex] 
$1 - a_{i1} - a_{i2}$ & $0.0004$ & $0.0010$ & $0.0016$ & $0.0026$ & $0.0031$ & $0.0120$ \\ 
\hline \\[-1.8ex] 
\end{tabular} 
\end{adjustbox}
\end{table} 
Now we discuss the model fits. 
We consider the following special case of the general model proposed in equation (\ref{eqn:og_model}):
\begin{align*}
   \mbox{Level 1:} \quad &y_{igk} | \theta_{igk} \stackrel{ind}{\sim} \text{Bernoulli}(\theta_{igk}),\\
    % \mbox{Level 2:} \quad & \phi(\theta_{igk} )= x'_{g(1)}\alpha+x'_{i(2)}\beta+ x'_{igk(3)}\xi_{c} + v_i+\lambda h(w_{igk}), \\
    \mbox{Level 2:} \quad & \mbox{logit} (\theta_{igk} )=  x'_{g(1)}\alpha+x'_{i(2)}\beta+ x'_{igk(3)}\xi_{c} + \lambda h(w_{igk})+v_i+u_{ig},\\
    \mbox{Level 3:} \quad & v_i \stackrel{iid}{\sim} N(0, \sigma^2_v),\;u_{ig}\stackrel{iid}{\sim} N(0, \sigma^2_u).
    %\mbox{Level 4:} \quad & \mbox{weakly informative priors on} \ \beta,  \mbox{and}\  \sigma_v,
\end{align*}
% I think the above model will reduce the bias problem (HB will be bouncing around the national estimate, unlike what we have now).
% This model, unlike the one you tried last time, is identifiable and so you should not experience problem in MCMC. 
To fit the models in Table \ref{tab: model_list}, we used the probabilistic programming language $\mathbf{Stan}$ \cite{stan2024} with the Statistical software $\mathbf{R}$ \cite{R}. The $\mathbf{R}$-package \texttt{rstan} \cite{rstan} provides the $\mathbf{R}$ interface to $\mathbf{Stan}$. 
% Our focus of this work was not to determine how different priors would affect our model fits or which exact prior specification will be better than others, thus 
For all intercepts (one for each Race × Ethnicity combination) and the slopes, we have used iid $N(0, 5^2)$ prior. For the variance components–$\sigma_v$ and $\sigma_u$, we have used the $\text{Cauchy}^+(0, 5)$ prior, truncated to the left at 0. Note that, such priors on the model hyperparameters are often called `weakly informative priors', or WIP in literature \cite{gelman2013bayesian}. These prior choices made sure that the posterior density will be proper. We ran each model for $4,000$ iterations, and the first $2,000$ were discarded as burn-in samples. Thus, all the posterior analyses were based on $2,000$ post-warm-up MCMC draws.

\subsection{Model selection}

After fitting all the 4 models, we have used a Model Selection criterion to select one `best' working model. The $\textbf{R}$-package \texttt{loo} \cite{loo} provides a quick and convenient way to diagnose and compare model fits, and we have used the same in this work. The software computes approximate Leave-One-Out Cross-Validation (LOO-CV) using Pareto smoothed importance sampling (PSIS), a relatively new procedure for regularizing importance weights. We refer to \cite{vehtari2017practical} for a detailed discussion on `loo' with Stan. As a byproduct of the calculations, \texttt{loo} can also obtain approximate standard errors for estimated predictive errors and for comparing predictive errors between two models. When comparing any two fitted models, the package can estimate the difference in their expected predictive accuracy by the difference in their $\hat{\text{elpd}}_{\text{loo}}$, the Bayes Leave-One-Out estimate of the expected log pointwise predictive density (elpd). This difference will be referred to as \texttt{elpd\_diff}. For a large enough sample size (used to fit the data) this difference approximately follows a standard normal distribution. This will facilitate all the model comparisons that we have used in this paper. The following Table \ref{loo_table} provides the LOO-CV comparison results for the 4 models.

\begin{table}[h] \centering 
  \caption{loo-cv comparison for the 4 models. Model 3 is selected as the best model.} 
  \label{loo_table} 
\begin{tabular}{@{\extracolsep{5pt}} ccccccccc} 
\\[-1.8ex]\hline 
\hline \\[-1.8ex] 
 & elpd\_diff & se\_diff \\ 
\hline \\[-1.8ex] 
M3 & $0$ & $0$ \\ 
M2 & -$0.958$ & $0.804$ \\ 
M1 & -$1.672$ & $0.836$ \\ 
M4 & -$64.216$ & $18.477$ \\ 
\hline \\[-1.8ex] 
\end{tabular} 
\end{table} 
From Table \ref{loo_table}, we see that LOO criterion selects the Model 3 as the best model, and it is better than the Model 2 as seen by the \texttt{elpd\_diff} which is significant at $\sim$68\%. Note that the survey weights enter linearly in the model M3. Following \cite{verret2015model}, we wanted to see how M3 would perform when the survey weights are entered in the model non-linearly, keeping everything else unchanged, viz.,  $\lambda\ln(w_{igk})$ and $\lambda w_{igk}^{-1}$ instead of $\lambda w_{igk}$. Naturally, to check the performance of these two new competing models, we again employed the LOO-criterion. From Table \ref{new-loo} we observe that neither log-transformed survey weights nor the inverse of the survey weights bring about any improvement to the already existing M3. So, we keep M3 as our best working model and follow-up with subsequent analyses with the Model 3 (M3).

\begin{table}[!htbp] \centering 
  \caption{loo-cv comparison of Model 3 varying weight functions} 
  \label{new-loo} 
\begin{tabular}{@{\extracolsep{5pt}} ccc} 
\\[-1.8ex]\hline 
\hline \\[-1.8ex] 
 & elpd\_diff & se\_diff \\ 
\hline \\[-1.8ex] 
M3 & $0$ & $0$ \\ 
M3(ln($w_{igk}$)) & -$0.329$ & $1.874$ \\ 
M3($w_{igk}^{-1}$) & -$3.644$ & $2.576$ \\ 
\hline \\[-1.8ex] 
\end{tabular} 
\end{table} 

Looking at the posterior analyses from the Model 3, outlined in Table \ref{tab6}, we see that the variable co-morbidity from the Facebook survey seem to be insignificant at 70\%. That is why we decided to run another model comparison between our chosen Model 3 and Model 3 without that particular variable. LOO analysis described in Table \ref{with-without-FB} suggests that removing the co-morbidity from the model does not bring about any significant difference. So we decided to leave that variable in our model.

\begin{table}[!htbp] \centering 
  \caption{loo-cv comparison of Model 3 and Model 3 without the co-morbidity rate from Facebook-CMU-Delphi COVID-19 Trends and Impact Survey} 
  \label{with-without-FB} 
\begin{tabular}{@{\extracolsep{5pt}} ccc} 
\\[-1.8ex]\hline 
\hline \\[-1.8ex] 
 & elpd\_diff & se\_diff \\ 
\hline \\[-1.8ex] 
M3 & $0$ & $0$ \\ 
M3 (without co-morb from FB) & -$0.740$ & $0.747$ \\
\hline \\[-1.8ex] 
\end{tabular} 
\end{table}

\subsection{Empirical justification for the bias correction approach}
\label{sec:empirical_justification}

\begin{figure}[h]
\centering
\includegraphics[width=0.6\linewidth]{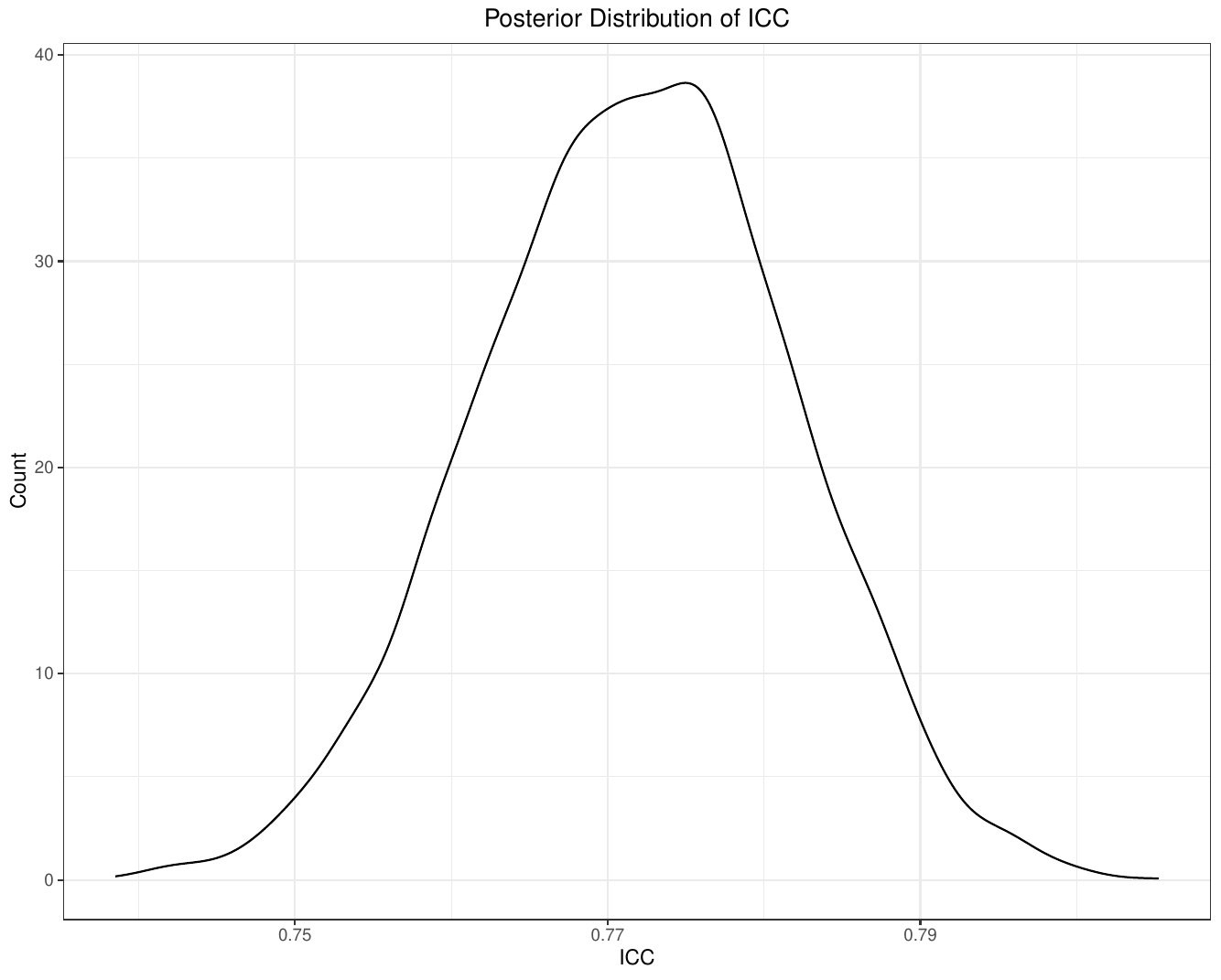}
\caption{Posterior distribution of intraclass correlation (ICC) for survey weights across post-stratification cells}
\label{fig:icc_weights}
\end{figure}

To empirically justify our bias correction approach described in Section 2.4, we conducted an analysis examining the structure of survey weights across post-stratification cells. We examined the intraclass correlation ($\rho$) for survey weights through a hierarchical Bayesian model. For survey weights (denoted as $w_{igk}$), we specified a hierarchical model:
\begin{align*}
w_{igk} &\sim N(\mu_{igk}, \sigma^2_e)\\
\mu_{igk} &= \beta_0 + v_{ig}\\
v_{ig} &\sim N(0, \sigma^2_v)
\end{align*} With weakly informative priors: $\beta_0 \sim N(0, 5^2)$, $\sigma_v \sim \text{Cauchy}^+(0, 5)$, and $\sigma_e \sim \text{Cauchy}^+(0, 5)$. The intraclass correlation coefficient was calculated as $\rho_w = \sigma^2_v/(\sigma^2_v + \sigma^2_e)$. Figure \ref{fig:icc_weights} shows the posterior distribution of the ICC for survey weights.

 The posterior distribution of $\rho$ was concentrated near 0.77, with $P( \rho > 0.75) = 0.985$. This high ICC value confirms strong within-cell homogeneity of weights, providing empirical evidence that supports our methodological approach. This demonstrates that our post-stratification structure effectively captures the structure of survey weights, with weights being highly homogeneous within cells (ICC $\approx 0.77$). This high within-cell homogeneity supports the approximation in our Hájek-type estimators. It is important to note that while we could conduct this empirical analysis for the observed survey weights, a similar analysis for the unobserved $\theta_{igk}$ values is not possible since these are latent quantities that cannot be directly measured or verified. Any remaining heterogeneity is addressed by the bias correction procedure described in Section 2.4, which provides additional protection particularly for cells in $\mathcal{G}_{2i}$ that lack direct representation in their respective areas.

\subsection{Findings from the selected model}
In the last subsection, we decided on one model that best fits our data with the help of LOO criterion using $\mathbf{Stan}$, \texttt{rstan} and \texttt{loo}. Now we present important findings from the selected `best' working model, Model 3. 

\begin{table}[!htbp] \centering 
  \caption{Posterior summary statistics for the model parameters of Model 3 -- the best Model chosen by LOO criterion} 
  \label{tab6} 
  \begin{adjustbox}{width=0.73\textwidth}
\begin{tabular}{@{\extracolsep{5pt}} ccccccccc} 
\\[-1.8ex]\hline 
\hline \\[-1.8ex] 
 & mean &  sd & 10\% & 15\% & 85\% & 90\% & $N_{eff}$ & $\hat{R}$ \\ 
\hline \\[-1.8ex] 
$\alpha_1$ & -$2.457$ & $0.888$ & -$3.608$ & -$3.361$ & -$1.566$ & -$1.349$ & $199.651$ & $1.002$ \\ 
$\alpha_2$ & -$1.232$ & $0.870$ & -$2.345$ & -$2.132$ & -$0.357$ & -$0.141$ & $201.311$ & $1.003$ \\ 
$\alpha_3$ & -$0.916$  & $0.878$ & -$2.078$ & -$1.843$ & -$0.034$ & $0.159$ & $187.865$ & $1.001$ \\ 
$\alpha_4$ & -$1.422$  & $0.873$ & -$2.540$ & -$2.347$ & -$0.545$ & -$0.316$ & $210.060$ & $1.003$ \\ 
$\alpha_5$ & -$1.391$  & $0.885$ & -$2.522$ & -$2.326$ & -$0.500$ & -$0.294$ & $198.465$ & $1.002$ \\ 
$\alpha_6$ & -$1.997$  & $0.896$ & -$3.157$ & -$2.932$ & -$1.096$ & -$0.905$ & $221.261$ & $1.002$ \\ 
$\alpha_7$ & -$1.121$  & $0.979$ & -$2.385$ & -$2.140$ & -$0.113$ & $0.116$ & $245.541$ & $1.002$ \\ 
$\alpha_8$ & -$1.574$ & $1.153$ & -$3.044$ & -$2.762$ & -$0.368$ & -$0.154$ & $302.163$ & $1.002$ \\ 
\text{co-morb} & -$0.573$ & $0.682$ & -$1.439$ & -$1.273$ & $0.109$ & $0.280$ & $202.383$ & $1.002$ \\ 
\text{survey wts} & -$0.112$  & $0.040$ & -$0.161$ & -$0.153$ &-$0.070$ & -$0.060$ & $1,040.340$ & $1.000$ \\ 
$\xi_{male}$ & $0.209$  & $0.019$ & $0.184$ & $0.189$ & $0.229$ & $0.233$ & $1,122.518$ & $1.001$ \\ 
$\xi_{female}$ & $0.118$ & $0.020$ & $0.092$ & $0.096$ & $0.139$ & $0.144$ & $952.271$ & $1.001$ \\ 
\text{rep \%} & -$0.833$ & $0.575$ & -$1.571$ & -$1.427$ & -$0.248$ & -$0.103$ & $385.690$ & $1.001$ \\
$\sigma_v$ & $0.121$ & $0.072$ & $0.026$ & $0.035$ & $0.196$ & $0.215$ & $31.222$ & $1.039$ \\ 
$\sigma_u$ & $0.162$ & $0.048$ & $0.108$ & $0.116$ & $0.212$ & $0.228$ & $29.895$ & $1.059$ \\ 
\hline \\[-1.8ex] 

\end{tabular} 
\end{adjustbox}
\end{table}

Table \ref{tab6} gives the posterior summary statistics for the selected best model, Model 3. The `mean' column gives the posterior means, the `sd' column gives the posterior standard deviations (estimated standard errors), the next 4 columns give the 4 percentiles, the column $N_{\text{eff}}$ gives the Effective Sample Sizes (ESS, $N_{\text{eff}}$) -- the number of independent samples that was used to replace the total $N$ dependent MCMC draws having the same estimation power as the $N$ autocorrelated samples, and finally the column $\hat{R}$ gives a measure of chain-equilibrium -- a value near 1 (but $< 1.1$) means that the chains have mixed well, and the posterior samples can be used with confidence for the posterior analyses (\cite{StanManual}, \cite{gelman2013bayesian}). As a default rule,  \cite{gelman2013bayesian} recommend running the simulation until $N_{\text{eff}}$ is at least $5m$, where $m$ is twice the number of independent chains/sequences. For our calculations, we ran 2 independent chains, each for $2,000$ iterations. This makes  $5m = 5 \times 2 \times 2 = 20$ for our case. As we can see from table \ref{tab6}, all the $N_{\text{eff}} > 20$, and $\hat{R} < 1.1$. Thus, we can use these posterior analyses with fair confidence, that the chains attained stationarity, and as well as achieved proper mixing. We use the posterior estimates to calculate our state-wise parameter of interest $\bar{Y}_i$ for all the 50 states and DC -- this essentially gives us the state-wise vaccine hesitancy estimates. These numbers along with their corresponding estimated model-based standard errors are displayed in Table \ref{tab7}. For easy comparison, the direct survey based estimates are also given in the same table. We note that that our synthetic HB methodology has smoothed the direct design-based estimates considerably. This is a well-known phenomenon in the literature \cite{Rao2015}.

\begin{figure}[]
    \centering
    \includegraphics[width=0.87\textwidth]{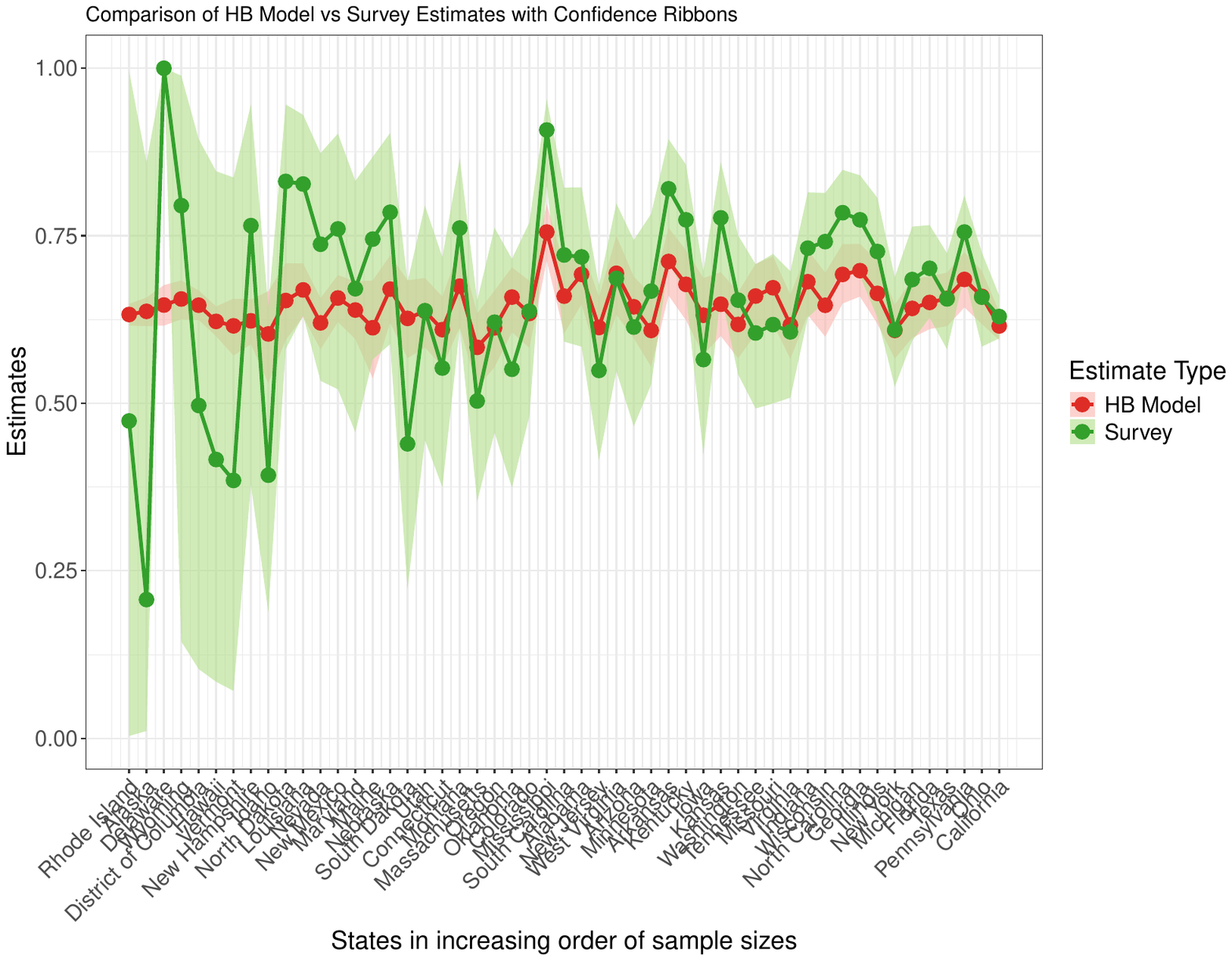}
    \caption{In x-axis, we have states in increasing order of UAS sample sizes. In y-axis, we have the point estimates of proportion of people who are ``hesitant" to get vaccinated from UAS survey (in green) and estimates from our HB Model -- Model 3 (in red) \textit{without} any bias correction as proposed in section 2.3. 95\% Credible Intervals for the HB model and 95\% Confidence Intervals for the direct estimates are shown in ribbons.}
    \label{fig:woBiasCorrection}
\end{figure}

\begin{figure}[]
    \centering
    \includegraphics[width=0.87\textwidth]{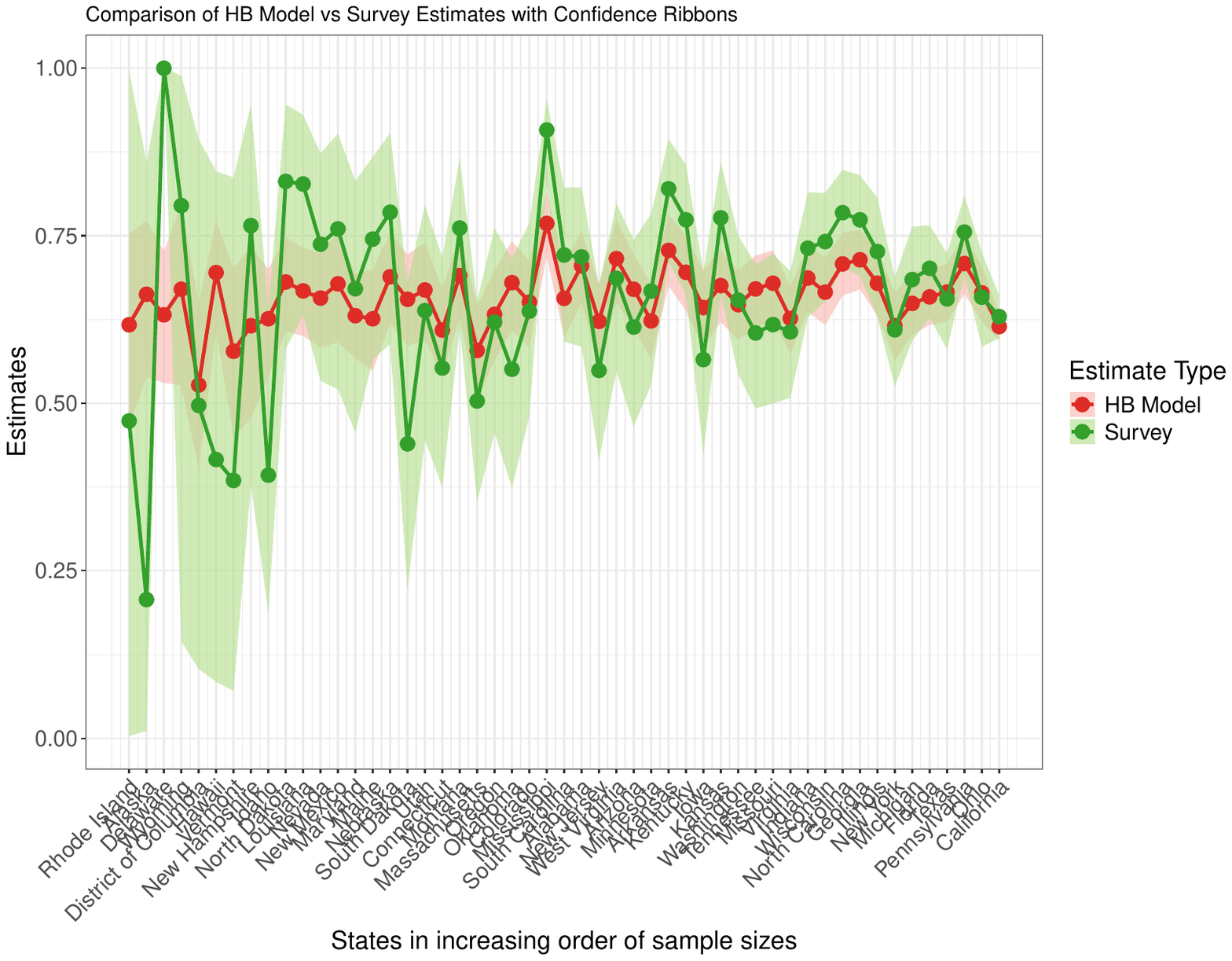}
    \caption{In x-axis, we have states in increasing order of UAS sample sizes. In y-axis, we have the point estimates of proportion of people who are ``hesitant" to get vaccinated from UAS survey (in green) and estimates from our HB Model -- Model 3 (in red) \textit{with} bias correction as proposed in section 2.3. 95\% Credible Intervals for the HB model and 95\% Confidence Intervals for the direct estimates are shown in ribbons.}
    \label{fig:wBiasCorrection}
\end{figure}

Next, we compare our HB model-based state-wise estimates with the UAS-survey-weighted state-wise estimates. From Figures \ref{fig:woBiasCorrection} and \ref{fig:wBiasCorrection}, we observe that while state-level UAS estimates with adequate sample sizes are generally reliable, estimates for states with small samples (e.g., Alaska, Delaware, Rhode Island, Vermont) show unreasonable values with extremely high design-based standard errors. In contrast, our HB method produces stable estimates even for states with minimal or no sample representation. Figure \ref{fig:woBiasCorrection} displays results without the bias correction procedure described in Section 2.3, while Figure \ref{fig:wBiasCorrection} shows results incorporating bias correction. The corresponding 95\% credible intervals for the HB estimates are also provided in both plots. While the point estimates remain similar with and without bias correction, the confidence intervals for the bias-corrected estimates better reflect the additional uncertainty inherent in the estimation process for areas with limited direct sample information.

% Next, we compare our HB model-based state-wise estimates with the UAS-survey-weighted state-wise estimates. \SD{Sir, I think we should add some comments about how HB model with bias correction have wider confidence ribbons, even though the point estimates are very close.} We plot the point estimates in the same graph. From the Figure \ref{fig:woBiasCorrection} and \ref{fig:wBiasCorrection} we observe that even though state-level UAS estimates for which enough sample are available are reliable, some of them, for which little to no samples were available in the UAS data, are unreasonably low/high due to the low sample sizes. Compared to that, our HB method produced stable estimates even for states with small or no sample. The corresponding 95\% credible intervals of the HB estimates are also given in the same plot. 

% \begin{figure}[]
%     \centering
%     \includegraphics[width=0.79\textwidth]{hes_HB_CI_new.pdf}
%     \caption{In x-axis, we have states in increasing order of UAS sample sizes. In y-axis, we have the point estimates of proportion of people who are ``hesitant" to get vaccinated from UAS survey (in blue) and estimates from our HB Model -- Model 3 (in red). The vertical lines around every HB estimates are their corresponding 95\% Credible Intervals.}
%     \label{fig:hes_HB_CI}
% \end{figure}

\begin{figure}[]
    \centering
    \includegraphics[width=0.73\textwidth]{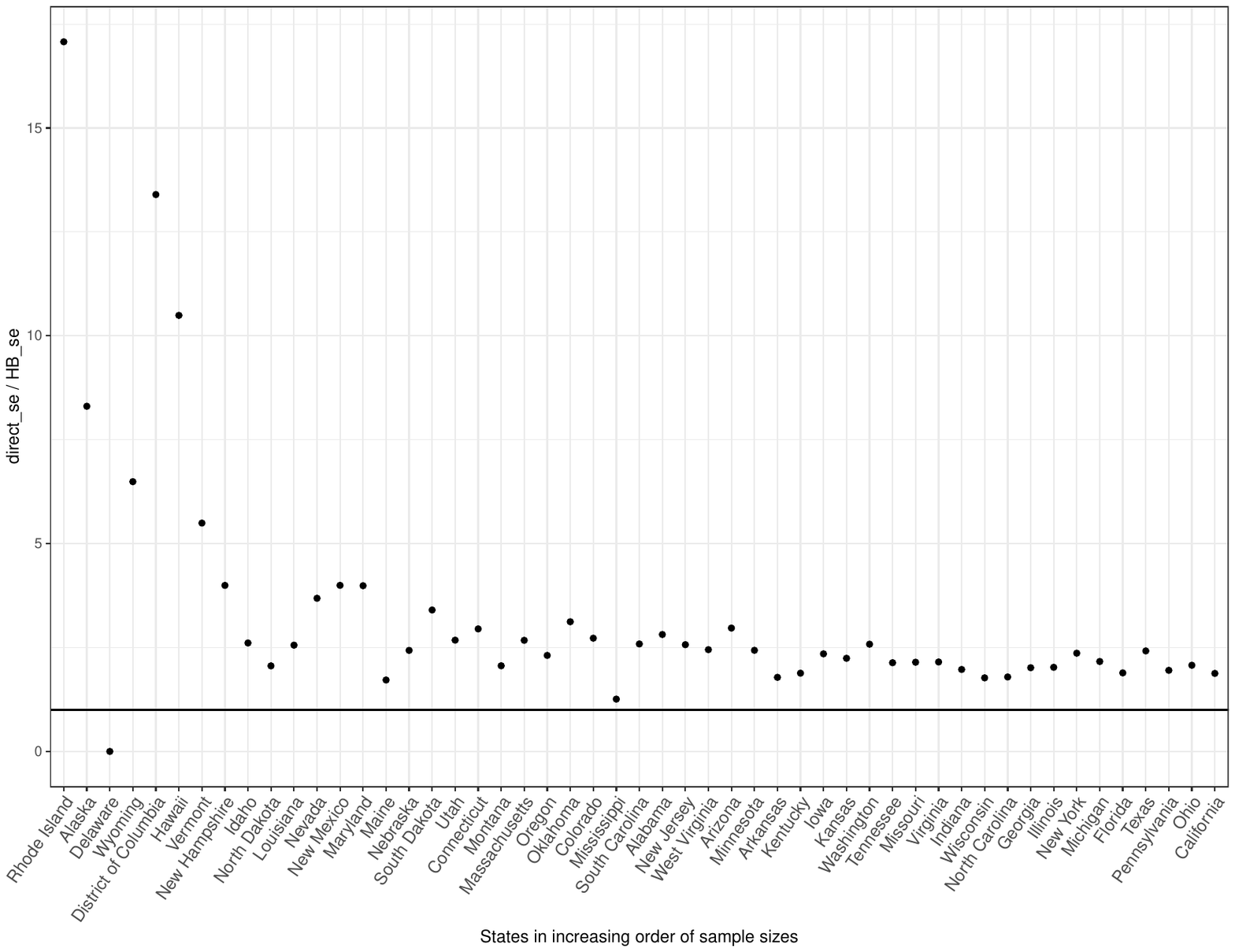}
    \caption{In x-axis, we have states in increasing order of UAS sample sizes. In y-axis, we have ratios of standard errors of direct UAS-survey based estimates to our HB model estimates. The horizontal dark line intersects the y-axis at 1.}
    \label{fig:se_ratios}
\end{figure}

From the Figure \ref{fig:se_ratios}, we can see how much improvement our HB estimates bring in over the direct UAS-survey based estimates in terms of estimated standard errors of the hesitancy estimates. The largest improvements came for the states that had lower sample sizes in the UAS survey data. We can still observe smaller standard error estimates even for the states with larger sample sizes.

\begin{table}[] 
\centering 
  \caption{HB hesitancy estimates (with and without bias correction) and direct UAS design-based estimates along with their estimated standard errors for all the 50 states and Washington D.C.} 
  \label{tab7} 
  
\begin{adjustbox}{width=\textwidth}
\begin{tabular}{lcrrrr|lcrrrr} 
\\[-1.8ex]\hline 
\hline \\[-1.8ex] 
\multicolumn{1}{c}{} & \multicolumn{1}{c}{} & \multicolumn{1}{c}{} & \multicolumn{3}{c}{Hesitancy Estimates (SE)} & \multicolumn{1}{c}{} & \multicolumn{1}{c}{} & \multicolumn{1}{c}{} & \multicolumn{3}{c}{Hesitancy Estimates (SE)} \\ 
\multicolumn{1}{c}{\#} & \multicolumn{1}{c}{State} & \multicolumn{1}{c}{n} & No Bias Corr. & With Bias Corr. & Direct & \multicolumn{1}{c}{\#} & \multicolumn{1}{c}{State} & \multicolumn{1}{c}{n} & No Bias Corr. & With Bias Corr. & Direct \\ 
\hline \\[-1.8ex] 
1 & Alabama & 73 & 0.721 (0.021) & 0.705 (0.028) & 0.718 (0.060) & 26 & Missouri & 117 & 0.686 (0.027) & 0.679 (0.027) & 0.617 (0.057) \\ 
2 & Alaska & 5 & 0.611 (0.022) & 0.663 (0.060) & 0.207 (0.187) & 27 & Montana & 52 & 0.689 (0.031) & 0.691 (0.032) & 0.762 (0.063) \\ 
3 & Arizona & 82 & 0.668 (0.024) & 0.670 (0.028) & 0.614 (0.072) & 28 & Nebraska & 37 & 0.643 (0.032) & 0.689 (0.035) & 0.785 (0.078) \\ 
4 & Arkansas & 93 & 0.752 (0.026) & 0.728 (0.027) & 0.820 (0.046) & 29 & Nevada & 28 & 0.615 (0.023) & 0.657 (0.036) & 0.737 (0.085) \\ 
5 & California & 1879 & 0.601 (0.009) & 0.615 (0.011) & 0.629 (0.016) & 30 & New Hampshire & 9 & 0.576 (0.033) & 0.616 (0.066) & 0.765 (0.131) \\ 
6 & Colorado & 60 & 0.631 (0.027) & 0.651 (0.033) & 0.638 (0.074) & 31 & New Jersey & 77 & 0.583 (0.026) & 0.622 (0.030) & 0.549 (0.067) \\ 
7 & Connecticut & 41 & 0.566 (0.030) & 0.609 (0.033) & 0.553 (0.088) & 32 & New Mexico & 33 & 0.669 (0.024) & 0.678 (0.042) & 0.760 (0.096) \\ 
8 & Delaware & 5 & 0.751 (0.026) & 0.632 (0.051) & 1.000 (0.000) & 33 & New York & 178 & 0.628 (0.018) & 0.616 (0.026) & 0.610 (0.042) \\ 
9 & DC & 7 & 0.701 (0.016) & 0.527 (0.062) & 0.497 (0.219) & 34 & North Carolina & 157 & 0.719 (0.021) & 0.708 (0.024) & 0.784 (0.037) \\ 
10 & Florida & 241 & 0.662 (0.019) & 0.659 (0.021) & 0.701 (0.035) & 35 & North Dakota & 23 & 0.669 (0.042) & 0.681 (0.036) & 0.831 (0.086) \\ 
11 & Georgia & 164 & 0.735 (0.019) & 0.714 (0.022) & 0.774 (0.038) & 36 & Ohio & 281 & 0.689 (0.017) & 0.665 (0.021) & 0.658 (0.036) \\ 
12 & Hawaii & 7 & 0.573 (0.019) & 0.695 (0.042) & 0.416 (0.203) & 37 & Oklahoma & 58 & 0.626 (0.028) & 0.680 (0.034) & 0.551 (0.089) \\ 
13 & Idaho & 23 & 0.506 (0.045) & 0.626 (0.041) & 0.393 (0.118) & 38 & Oregon & 56 & 0.551 (0.034) & 0.633 (0.034) & 0.621 (0.079) \\ 
14 & Illinois & 176 & 0.674 (0.022) & 0.679 (0.026) & 0.726 (0.045) & 39 & Pennsylvania & 255 & 0.733 (0.015) & 0.709 (0.023) & 0.756 (0.030) \\ 
15 & Indiana & 152 & 0.687 (0.024) & 0.687 (0.025) & 0.731 (0.048) & 40 & Rhode Island & 3 & 0.594 (0.019) & 0.617 (0.075) & 0.474 (0.321) \\ 
16 & Iowa & 94 & 0.602 (0.030) & 0.643 (0.029) & 0.565 (0.071) & 41 & South Carolina & 71 & 0.701 (0.023) & 0.657 (0.028) & 0.721 (0.058) \\ 
17 & Kansas & 107 & 0.676 (0.022) & 0.676 (0.028) & 0.777 (0.050) & 42 & South Dakota & 39 & 0.543 (0.036) & 0.655 (0.035) & 0.440 (0.122) \\ 
18 & Kentucky & 93 & 0.680 (0.026) & 0.696 (0.029) & 0.774 (0.049) & 43 & Tennessee & 114 & 0.622 (0.026) & 0.671 (0.028) & 0.605 (0.055) \\ 
19 & Louisiana & 25 & 0.721 (0.028) & 0.668 (0.034) & 0.827 (0.071) & 44 & Texas & 245 & 0.661 (0.015) & 0.666 (0.022) & 0.656 (0.037) \\ 
20 & Maine & 36 & 0.591 (0.044) & 0.626 (0.038) & 0.745 (0.076) & 45 & Utah & 40 & 0.630 (0.034) & 0.669 (0.037) & 0.638 (0.090) \\ 
21 & Maryland & 34 & 0.670 (0.024) & 0.631 (0.032) & 0.671 (0.096) & 46 & Vermont & 7 & 0.558 (0.037) & 0.578 (0.066) & 0.385 (0.204) \\ 
22 & Massachusetts & 56 & 0.531 (0.029) & 0.579 (0.036) & 0.504 (0.077) & 47 & Virginia & 151 & 0.602 (0.022) & 0.627 (0.026) & 0.607 (0.048) \\ 
23 & Michigan & 192 & 0.701 (0.020) & 0.649 (0.024) & 0.685 (0.043) & 48 & Washington & 110 & 0.613 (0.021) & 0.647 (0.027) & 0.654 (0.053) \\ 
24 & Minnesota & 83 & 0.674 (0.027) & 0.623 (0.028) & 0.667 (0.065) & 49 & West Virginia & 81 & 0.671 (0.026) & 0.716 (0.032) & 0.686 (0.064) \\ 
25 & Mississippi & 60 & 0.769 (0.024) & 0.768 (0.024) & 0.908 (0.031) & 50 & Wisconsin & 155 & 0.644 (0.023) & 0.666 (0.026) & 0.741 (0.041) \\ 
 &  &  &  &  &  & 51 & Wyoming & 5 & 0.659 (0.028) & 0.670 (0.070) & 0.795 (0.184) \\ 
\hline \\[-1.8ex] 
\multicolumn{11}{l}{\scriptsize{Note: n = sample size from UAS, ``SE" = Standard Error (estimated), ``DC" = District of Columbia. All estimates are shown with their standard errors in parentheses.}}\\
\end{tabular} 
\end{adjustbox}
\end{table}

Table~\ref{tab7} presents state-level vaccine hesitancy estimates using three different approaches: hierarchical Bayes without bias correction, hierarchical Bayes with bias correction (as described in Section~2.3), and direct UAS design-based estimates. For each estimate, the corresponding standard error is shown in parentheses. The comparison reveals several important patterns. First, bias correction substantially impacts states with small sample sizes, such as Delaware ($n=5$, 0.751 $\to$ 0.632), District of Columbia ($n=7$, 0.701 $\to$ 0.527), and Hawaii ($n=7$, 0.573 $\to$ 0.695). Second, states with minimal samples show larger increases in standard errors after bias correction, appropriately reflecting the additional uncertainty inherent in synthetic estimation (also see Figure \ref{fig:wBiasCorrection}). Third, both HB approaches provide substantially more stable estimates than direct survey estimates for states with small samples, as evidenced by the extremely high standard errors for direct estimates in states like Rhode Island (SE = 0.321), Vermont (SE = 0.204), and Hawaii (SE = 0.203). The table highlights particularly dramatic improvements for Alaska, where the direct estimate of 0.207 with SE = 0.187 is replaced by a more plausible bias-corrected estimate of 0.663 with SE = 0.060, and for Delaware, where the implausible direct estimate of 1.000 with SE = 0.000 is replaced by 0.632 with SE = 0.051.

\section{Concluding remarks}

In conclusion, this paper introduces a novel hierarchical Bayes estimation procedure for finite population means in small areas, employing a sophisticated data linkage technique that integrates a primary (probability) survey with diverse data sources, including non-probability survey data from social media. In our elaborate model, assumed only for the sample respondents, we include weights from the primary survey and factors potentially influencing the outcome variable of interest in order to reduce informativeness of the sample. If the applied data analysis suggests that the weights are significant in the model, then this will imply that the sampling is informative. One should no longer assume the validity of the same model for the respondents and the population. Unlike many existing works, we acknowledge and address this critical step, assuming a simple model for the finite population regardless of the informativeness of sampling. The resulting hierarchical synthetic estimate is designed with the goal of reduced sensitivity to model misspecifications, with less reliance on assumed models compared to fully model-based approaches. This is achieved conceptually through our approach of only assuming an elaborate model for observed units while using a synthetic approach for unobserved units. While the full empirical evaluation of this theoretical benefit would require extensive simulation studies beyond the scope of this paper, our methodological design inherently provides some protection against model misspecification through its hybrid nature. Using the same notations as we did in section 2, this reduced dependence is demonstrated by ignoring the set of cells $\mathcal{G}_{3i}$, instead of using the assumed model to estimate the corresponding population mean $\Theta_{3i}$. Drawing inspiration from seminal works, our methodology is versatile and applicable to various finite population mean estimation challenges. However, it should be noted that our approach may fail to produce small area estimates in specific scenarios, requiring the validation of a crucial condition before proceeding with the data analysis. Again, using the same notations, for a probability survey data, it is essential that the condition $(1 - a_{1i} - a_{2i}) \approx 0$ is met. This condition needs to be checked before proceeding with the data analysis. Finally, we illustrate the application of our methodology in generating proportion estimates (small area means of a binary outcome variable) from a COVID-19 survey, statistically linking information from diverse data sources. Future extensions of this work may explore similar estimation procedures for jointly modeling categorical and continuous response variables. We are currently investigating this research direction.

\section*{Acknowledgements}

The research was partially supported by  the U.S. National Science Foundation Grant SES-1758808 awarded to the second author. The project described in this paper relies on data from survey(s) administered by the Understanding America Study (UAS), which is maintained by the Center for Economic and Social Research (CESR) at the University of Southern California (USC). The content of this paper is solely the responsibility of the authors and does not necessarily represent the official views of USC or UAS. The collection of the UAS COVID-19 tracking data is supported in part by the Bill \& Melinda Gates Foundation and by grant U01AG054580 from the National Institute on Aging. This research uses survey data from Carnegie Mellon University’s Delphi Group through data sharing. Please visit \url{https://cmu-delphi.github.io/delphi-epidata/symptom-survey/}, for further information on the survey.

\bibliographystyle{chicago}
 % \bibliographystyle{plainnat}
% \bibliography{arxiv_new.bbl}
\bibliography{mybib.bib, references.bib}

\end{document}